\documentclass{aastex}
\usepackage{emulateapj5}

\shorttitle{Barred Galaxies without Disks?}
\shortauthors{Gadotti \& de Souza}

\begin{document}

\title{NGC 4608 and NGC 5701: barred galaxies without disks?}

\author{D. A. Gadotti and R. E. de Souza}
\affil{Departamento de Astronomia, Instituto de Astronomia, Geof\'{\i}sica e Ci\^encias
Atmosf\'ericas, \\ Universidade de S\~ao Paulo \\ Rua do Mat\~ao, 1226 -- Cid. Univers.
CEP 05508-900, S\~ao Paulo -- SP, Brasil}
\email{dimitri@astro.iag.usp.br \\ ronaldo@astro.iag.usp.br}
\slugcomment{Scheduled to The Astrophysical Journal Letters February 1st issue}

\begin{abstract}
We have performed for the first time in the literature a 2D structural analysis on the barred
lenticular and face--on galaxies NGC 4608 and NGC 5701. The results indicate that these galaxies
either have never had large disks or their disks were almost completely destroyed by their strong
bars, due to secular evolution processes. We discuss these surprising conclusions checking for
signs of secular evolution, considering bar forming instabilities, and suggesting, based on N-body
simulations, a new mechanism to form bars in spheroids, which includes non-spherical halos.
Quantitative predictions from our new mechanism are compared with those from other recent models for
bar formation and evolution.
\end{abstract}


\section{Introduction}

The most widely accepted mechanism to produce bars in galaxies is based on a global instability in
cool stellar disks, where cool means rotationally supported (see, e.g., \citet{bin87}). However,
it was already argued by \citet{too81} that a high central density disk would not form a bar since
an Inner Lindblad Resonance (ILR) would stop the bar instability \citep{sel00}. There are
nevertheless several evidences for the presence of dense centres and  ILR's in galaxies, including barred ones
\citep{sel01}. Recently, \citet{sel99} have numerically confirmed that even dynamically cool disks
can not develop a bar if it has the observed high central densities. Thus, how galaxies form bars
remains a question to be properly answered. Possibilities are discussed in the papers by Sellwood cited
above, but all are easily rejected in the same papers.
The problem is even worse for lenticular galaxies, since these are
not cool stellar systems and generally speaking have the most massive bulges
among disk galaxies in the Hubble sequence.
But the existence of barred lenticulars is indisputable! For instance, in the Revised
Shapley--Ames Catalog \citep{san81}, $1/4$ of the lenticular galaxies brighter than 14 in the B
band are barred. A related issue concerns the role of halos in bar formation, which were once thought to
stabilize disks against the formation of a bar. Recent results, however, indicate that, while producing smaller
growth rates of bar-forming modes, halos can induce stronger bars
\citep[and references therein]{ath02}.
In this Letter, we will present results of a detailed structural analysis on two barred lenticulars, which
indicate that either their disks have almost completely disappeared, probably by secular evolution
processes, or that bars were formed without disks. Moreover, we present numerical simulations
which show how the later possibility can be attained with non-spherical halos.

\section{Structural Analysis}

The two galaxies discussed in this Letter were observed in optical (B, V, R, I) and near-infrared (Ks)
broadbands in the Steward Observatory 61" and Bok telescopes, respectively.
After standard reduction procedures, the images were used in a 2D structural analysis algorithm
developed by \citet{sou97}. The algorithm was constructed in order to model the surface
brightness profiles of galaxies using two components: a bulge obeying the
S\'ersic luminosity profile and an exponential disk.
Besides of being thoroughly tested in model galaxies, the algorithm has been already applied in a
sample of 39 Sbc galaxies \citep{gad99,anj03} and 51 ellipticals and lenticulars \citep{sou03}
providing valuable structural information.

The reader is referred to the above cited references for details on the algorithm.
Essentially, the galaxy images are transformed into a matrix in which each point represents a pixel
value. The user provides initial values for the structural parameters of bulge and disk (e.g., effective
surface brightness, ellipticity, S\'ersic index etc.) and the algorithm then tries to fit the galaxy image
with model images of bulge and disk, by varying their structural parameters, minimizing the $\chi^2$
deviation at each pixel until it reaches a convergence limit.
A total of 11 parameters are needed to fully describe the model and these are obtained from the
image fitting. Using the fitted parameters, one can build a
model galaxy and construct a residual image which can be useful in showing hidden sub-structures.

An important point to clarify now is that, in the case of strongly barred galaxies, like those studied here,
the code performs a decomposition in 2 components that may be thought as the bar/bulge and the disk.
The reader should keep this in mind when we refer to ``bulge'' hereafter.

NGC 4608 is catalogued as a SB0 galaxy in the Revised Shapley--Ames Catalog \citep{san81}
and by \citet{van90} with CCD images. Previous surface photometry analysis was done
by \citet{ben76}, with 1D fitting on shallow photographic plates,
and by \citet{woz91}, fitting ellipses to the isophotes on
CCD frames. Both studies concluded that this galaxy presents a low surface brightness disk.
NGC 5701 is described as having an intermediate morphological type between SB0 and SBa by a
number of authors (see, e.g., \citet{san81}).

When applied to NGC 4608 and NGC 5701 the algorithm retrieves a {\em very} low surface
brightness disk. The major component is indubitably represented by the bar/bulge. Thus
one is left to conclude that there is practically no disk present in these
galaxies. The presence of a large Freeman disk like the ones observed in spirals
can be readily discarded.
Figure 1 shows the 2 galaxies and their residual images after subtraction of the model images. The
direct images at left have an isophotal map overlayed with 0.5 mag interval
between each contour level. These images refer to the R broadband but similar results were
obtained in all the other bands. It is worth noting that not even in Ks a disk was found, which could
be a possibility if the disk was hidden by dust or contained mostly very old population stars. Our
surface photometry reaches the level of about 26 mag/arcsec$^2$ in B and 23 mag/arcsec$^2$ in Ks,
with comparable values reached in the other passbands. Thus our photometry is deep enough to indicate that
the absence of a disk is real. Even a faint disk would be detected. If there is a disk in these galaxies
it has to have a luminosity far lower than the one measured in normal S0's. Moreover, we do detect
disks in other galaxies with images taken during the same nights, with the same exposure times and
applying the same analysis.

Looking at the images at left in Fig. 1, one can already notice what seems to be a somewhat empty
region around the bar in both galaxies. These empty regions appear much clearer in the residual
images at right. In the case of NGC 4608 the residual image shows clearly the bar and the lens. For
NGC 5701, one is left with the bar, the outer ring and the inner ring, which is not visible in the
direct image. The important point to stress here is that, after the
subtraction of only a bulge model, there is no clear sign of the presence of a disk. We should
remark that these objects are genuine S0 galaxies as many authors have argued and
because the conspicuous bars {\em should} be a disk perturbation. Thus, one must face the
possibility of formation and maintenance of a bar in the almost total absence of a disk!

Another possibility is that secular evolutionary processes induced by bars in their host
galaxies \citep[and references therein]{gad01} have disturbed the structure of an originally
healthy normal disk. In principle, these strong bars could have transferred material
from the initial disk to the bulge. Then, what seems now a lens in NGC 4608 could be the outer
remains of the disk. The outer ring in NGC 5701 could also be the signature of a pre-existent disk.
For this galaxy too, there is a hint of inner disk remaining in the residual image.

To test the evolutionary hypothesis,  we may consider the question
whether the stars we see today in the bar once belonged to a disk. Could the luminosity we detect in
the bar make up a disk if distributed accordingly? To answer that, we have determined
the luminosity in all bands within a radius equal to the bar length (that was determined examining
the ellipticity and position angle radial profiles) for the direct and residual images. The difference
between them corresponds to the bulge luminosity, while the luminosity in the residual images
may be attributed to the assumed disk. Thus we calculated what would have been a bulge/disk
luminosity ratio for these galaxies before the secular evolution processes took place. This ratio is
$\approx 2$ and therefore it is compatible with the secular evolution hypothesis since the bulge/disk
ratio for normal S0's is around 2 \citep{bin98}.

While bar formation without disks seems to be a quite unusual possibility, it is not at all trivial that
secular processes can be so strong as to destroy a disk almost completely. We suggest, in
the next section, how bars could be formed in galaxies without disks.

\section{N-Body Simulations}

The idea of having a bar forming without a disk is very powerful given its simplicity. It is based on the
assumption that non-spherical dark matter halos should exist. As noted by \citet{fre88} model dark halos
are generally triaxial and may be prolate. Moreover, their axial ratios can be quite extreme,
reaching values around 3, while a ratio of 2 is common. Given that these halos are large and
massive, it is reasonable that they could exert a strong influence on the dynamics of a stellar
system which is embedded in such a halo. We performed numerical simulations to check if such a
configuration could make a spheroid turn into a bar, i.e., a triaxial (or prolate) eccentric structure.

Thus, we simulated the evolution of a Plummer sphere embedded in a rigid halo represented by a
logarithmic potential, which produces flat rotation curves and can be used to easily modify the halo
core radius, the mass within it and the halo ellipticity  \citep{bin87}. The Plummer sphere has a
characteristic radius of 2 kpc, extends to 17 kpc, and a total mass of $1.2 \times 10^{11} $M$_\sun$.
The halo parameters modified were the axial ratios (from 1 to 4, prolate and triaxial), the core radius
(from 6 kpc to 10 kpc) and the core mass (from $0.9 \times 10^{11} $M$_\sun$ to $1.2 \times 10^{11} $M$_\sun$).
These values for the core properties are usually found in the literature (e.g., \citet{beg91}).

The simulations were performed with the {\sc nemo} package \citep{teu95} with $10^5$ particles, using
\citet{bar86} algorithm. The softening parameter was $\simeq 0.03$, and the opening angle was $0.7$.
Energy and the center of mass were conserved better than 0.1\%, typically.

Several experiments were performed varying the halo parameters. We noticed that: {\bf (i)}, the sphere is stable
when the axial ratios are low, remaining approximately spherical for any typical values for the core radius
and mass; {\bf (ii)}, oval distortions, weak and strong bars are formed when raising the axial ratios from
$\approx 2$ to $\approx 4$, not depending substantially on the other core parameters; and {\bf (iii)}, these results
do not depend whether the halo is prolate or triaxial. Figure 2 shows a clarifying example. It shows how a Plummer
sphere can be transformed into a bar within the dynamical influence of a triaxial halo with axial ratio equals 3,
after 1 Gyr. Comparing Fig. 2 with the left panels of Fig. 1, one can see that the bar formed
in our simulations is a good representation of the bars in NGC 4608 and NGC 5701. Also, the size of the
bars in these galaxies and in our simulations is similar ($\approx$ 10 kpc).

Our new bar formation mechanism needs halos with axial ratios around 3 and the results from cosmological
simulations by \cite{fre88} indicate that at least some halos must have these high values for their axial ratios.
More recent cosmological simulations by, e.g., \cite{war92}, show that the axial ratios of halos have a wide
distribution going to values as high as 3, while the simulations of \cite{bul01} indicate a typical value of around
1.5, but with extreme values going as high as 5. Concerning observations, recent results indicate values ranging
from 1.25--2 (\citet{buo02,sac99}) to more than 3 (\citet{sac90}).

\section{Discussion and Conclusions}

There are thus two possible scenarios that could explain the existence of bars in systems (almost) devoid of discs.
The first one is that bars can form in diskless systems, as described in the previous section. The second one
is that such systems are extreme examples of the evolutionary scenario proposed recently by \citet{ath03}.
A weak bar forms initially, and grows by losing angular momentum to the external disk and the halo,
via resonant stars (see \citet{ath02,atm02}). In extreme cases it could ``consume'' all or most of the disk material,
so that the end product would be a bar in a halo, with very little, if any, disk left, i.e., what we have found for
NGC 4608 and NGC 5701. Thus, these galaxies can well be extreme examples of this evolutionary scenario.
With this mechanism, strong halos lead to strong bars, which is also the case.

Let us now evaluate the predictions of the Athanassoula's model (ATH) and our model of bar formation without
disk (GDS) and make a quantitative comparison with the results for NGC 4608 and NGC 5701. For that, snapshots
of the final result from both models were scaled in size and intensity to allow for a meaningful comparison with the
observed galaxies. Figure 3 shows radial profiles of the ellipticity of the two galaxies and of both
models. One sees that GDS provides a very good fit to the galaxies in most of the profile, while ATH fits reasonably
well only the last fifth part of it. Figure 4 presents intensity cuts along the major and minor axes of the bar in the
models and the galaxies. One can see that these cuts are better predicted by the ATH model. Another strong point
favouring the secular evolution hypothesis is the fact that a boxy--peanut shape develops in the ATH model
when viewed edge--on, as we know is the case for many observed bars (e.g., \citet{bur99}). This does
{\em not} occur in the GDS model.

Another powerful quantitative comparison of the face--on shapes may be performed if one calculates the Fourier
components of the intensity distribution projected onto the equatorial plane, like in, e.g., \citep{atm02}. Figure
5 shows the results of the Fourier analysis. An inclination angle of 15 degrees was assumed for both galaxies
to be deprojected before the components were calculated, in agreement with values found in the literature
(\citet{jun97}) and from analysing single dish HI velocity measurements (\citet{hay98}). The main difference
between the behaviour of the Fourier components in the ATH and GDS models is that there is a maximum in
the former but not in the later. Thus the data from NGC 4608 favours the ATH model. The position of the
maximum is not relevant here since it varies considerably from one simulation to another, depending on the
Q Toomre parameter (see, e.g., \citet{bin87}). The maximum is nearer to the center in dynamically cold
systems (Athanassoula, private communication). However, NGC 5701 does not seem to have a maximum,
which supports the GDS model, unless a maximum is found in a {\em very} deep image.
The reader must bear in mind that both models are in no way a specific fit to the galaxies under study,
but they show that extreme cases of barred galaxies with hardly any disc are possible.
It is clear that a definite conclusion should not be taken at this stage. More data (including spectroscopy)
are necessary on these two galaxies and S0's in general. On the other hand, simulations
may now be directed to address this question further.

Nonetheless, two not necessarily mutually exclusive conclusions can be
tentatively postulated: {\bf (i)} - bars can be formed in spheroids through the dynamical effects of a sufficiently
eccentric halo, without the need of a stellar disk, and {\bf (ii)} - secular evolution in barred galaxies can be strong
enough to almost completely destroy their disks.

The first possibility puts some hope in our struggle to understand how bars form in galaxies, since
other mechanisms have several serious drawbacks. Of course, this mechanism alone can not be responsible
for all the observed bars, since it needs very eccentric halos (and there are disk galaxies!),
but could be at least for some barred lenticulars.
Also, it not excludes the possibility of other bar forming instabilities acting together and thus can be explored
to overcome some of the difficulties. On the other hand, it is based on very simplistic numerical experiments.
More realistic simulations can be performed, e.g., including a live halo and gas, which can
then account for other observed sub-structures, like the rings in NGC 5701. However, the essential idea, we
showed now, seems to be correct. The second possible conclusion shows how strong secular evolutionary
processes may be and so how seriously they should be considered in models of galaxy formation and evolution.

\acknowledgments

We are indebted to the referee, Lia Athanassoula, for many suggestions and comments that helped to greatly
improve our work, and for letting us use her results prior to publication. It is a pleasure to thank Ron Buta and
Ivo Busko for help on deprojecting galaxy images and Fourier analysis. DAG would like to thank Peter Teuben for
invaluable help on {\sc nemo}. Financial support from FAPESP grant 99/07492-7 is acknowledged.

\epsscale{0.8}

\begin{figure}
\plotone{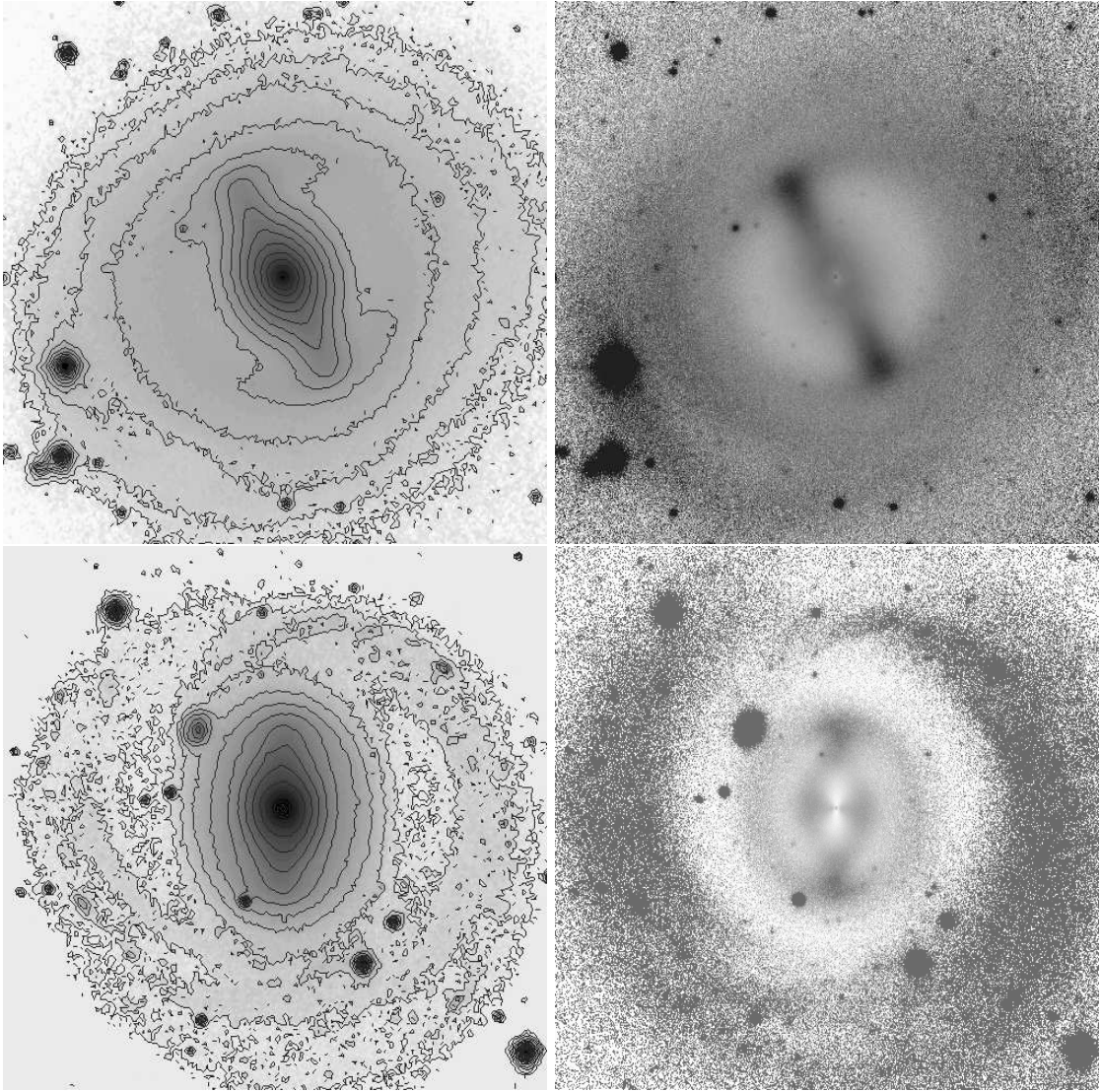}
\caption{Results from the structural analysis. Top panels refer to NGC 4608 while bottom
panels refer to NGC 5701. At left we present direct R images with an isophotal map. At right we show
residual images after the subtraction of a bulge model only. Note the absence of a disk.}
\end{figure}

\epsscale{0.3}

\vspace{1cm}

\begin{figure}
\plotone{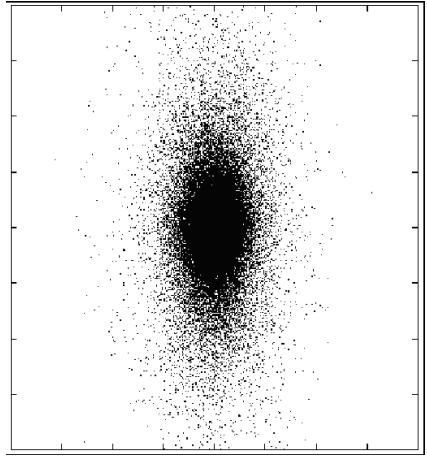}
\caption{Final structure formed from a Plummer sphere under the dynamical
influence of an eccentric halo after 1 Gyr. Compare with the bars in the left panels of Fig. 1.}
\end{figure}

\epsscale{0.4}

\begin{figure}
\plotone{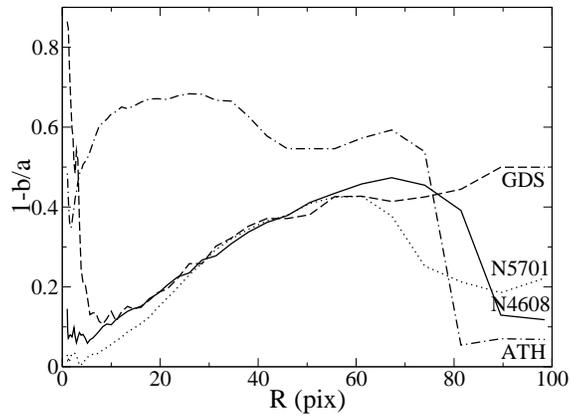}
\caption{Ellipticity radial profiles for NGC 4608 and NGC 5701, as well as for both the GDS
and ATH models.}
\end{figure}

\vspace{1cm}

\begin{figure}
\plotone{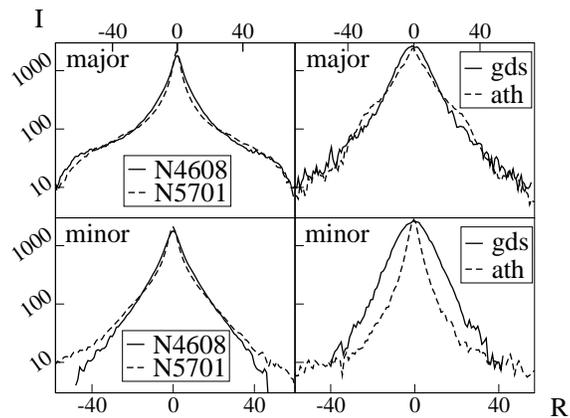}
\caption{Intensity profiles along the major and minor axes of the observed galaxies and the
models.}
\end{figure}

\epsscale{0.8}

\begin{figure}
\plotone{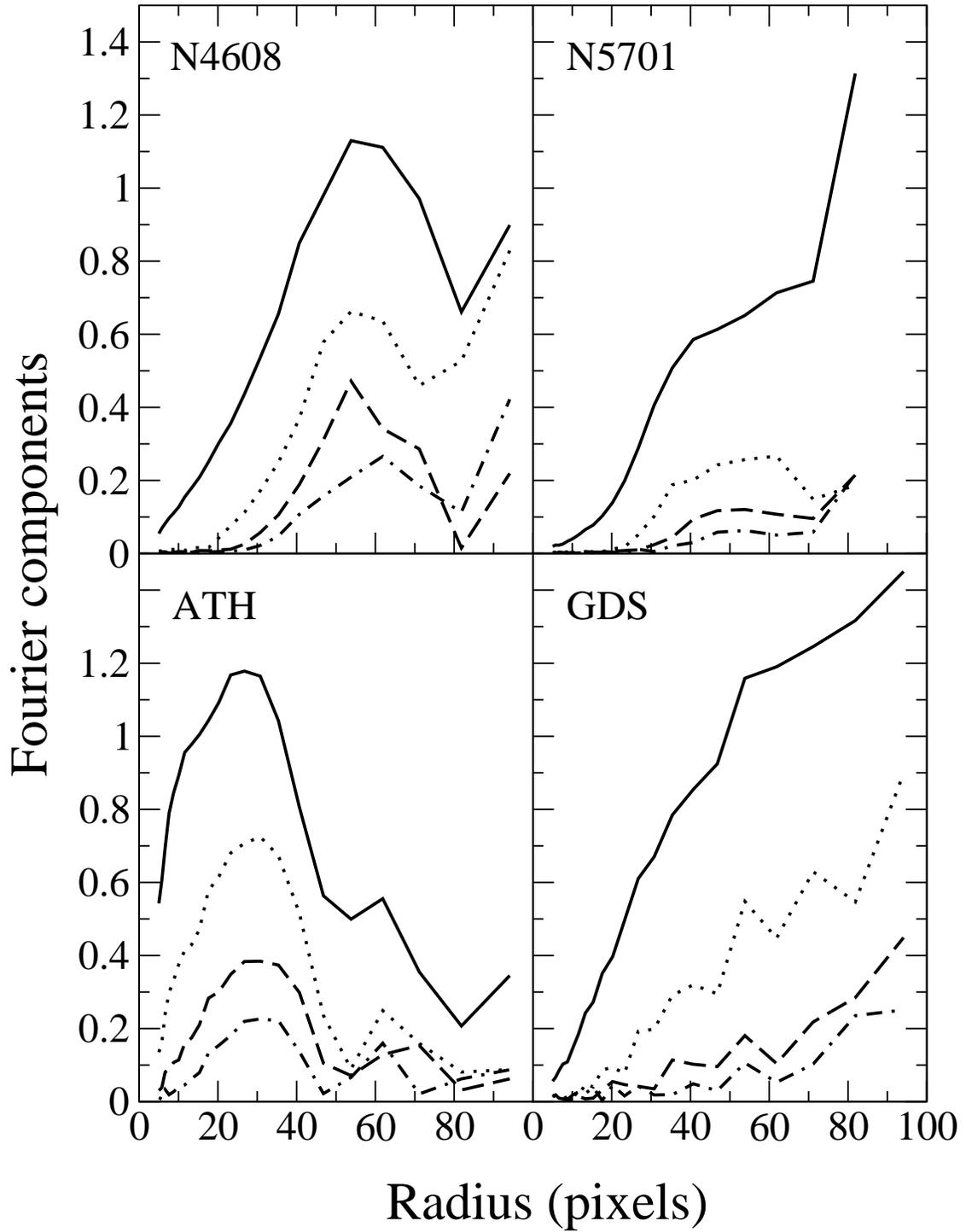}
\vspace{1cm}
\caption{Radial profiles of the Fourier components of the observed galaxies and the models.
The solid line refers to $m=2$, while the dotted line to $m=4$, the dashed line to $m=6$ and the
dash--dotted line to $m=8$.}
\end{figure}

\end{document}